\documentclass[%
    aps, prx,
    amsmath,amssymb,
    reprint,
]{revtex4-2}

\usepackage{bbm}
\usepackage{graphicx} % Include figure files
\usepackage{dcolumn} % Align table columns on decimal point
\usepackage{bm} % Bold math
\usepackage[T1]{fontenc}
\usepackage{mathptmx}
\usepackage{xcolor} % For defining a note style.

\RequirePackage{fix-cm} % Removes a font-warning.

\graphicspath{{gfx/}}

\renewcommand{\.}{\hspace{1pt}}

\newcommand{\1}{\mathbbm{1}}
\renewcommand{\i}{\mathchoice{}{\hspace{1pt}}{}{}\text{i}}
\renewcommand{\d}{\text{d}}

\newcommand{\ke}[1]{\vert #1 \rangle}

\let\bra\br
\let\ket\ke
\let\braket\ev
\newcommand*{\defeq}{\binoppenalty=\maxdimen\relpenalty=\maxdimen\mathrel{\vcenter{\baselineskip0.6ex \lineskiplimit0pt\hbox{.}\hbox{.}}}\hspace{0.8pt}=}

\begin{document}

    \title{The role of dephasing for dark state coupling in a molecular Tavis-Cummings model}

    \author{Eric Davidsson}
    \email{eric.davidsson@fysik.su.se}
    \author{Markus Kowalewski}%
    \email{markus.kowalewski@fysik.su.se}
    \affiliation{Department of Physics, Stockholm University, Albanova University Center, SE-106 91 Stockholm, Sweden}

    \date{\today}

    \begin{abstract}
        Collective coupling of an ensemble of particles to a light field
        is commonly described by the Tavis--Cummings model. This model includes
        numerous eigenstates which are optically decoupled from the optically bright
        polariton states. To access these dark states requires breaking the symmetry
        in the corresponding Hamiltonian. In this paper, we investigate the influence of non-unitary processes
        on the dark state dynamics in molecular Tavis--Cummings model.
        The system is modelled with a Lindblad equation that includes pure dephasing,
        as they would be caused by weak interactions with an environment,
        and photon decay.
        Our simulations show that the rate of the pure dephasing, as well as the
        number of particles, has a significant influence on the dark state population.

    \end{abstract}

    \maketitle

    \section{Introduction}

    Polaritonic chemistry---where
    chemical reactions are studied in the prescience of a strongly coupled electromagnetic field---is
    continuing to attract interest in fields ranging from chemistry to quantum optics,
    as shown by recent review studies
    \cite{Torma-Barnes-2014,Flick-etal-2017,Ribeiro-etal-2018,FriskKockum-etal-2019,Hertzog-etal-2019,Hirai-etal-2020,Fregoni-etal-2022}.
    These systems have been explored in what is becoming a rich landscape of different experiments.
    Examples includes demonstrating single-molecule strong coupling \cite{Chikkaraddy-2016},
    investigating vibrational strong coupling \cite{Simpkins-etal-2015,Damari-etal-2019,Thomas-etal-2019},
    strong coupling in J-aggregates \cite{Munkhbat-etal-2018},
    photo-isomerization \cite{Hutchison-2012,Mony-etal-2021},
    and two-dimensional spectroscopy of polaritonic systems \cite{Xiang-etal-2018,Mewes-2020}.
    Along with experiments, the community has also made significant theoretical advancements.
    Among the investigated phenomena, we can highlight
    studies of conical intersections \cite{Ulusoy-etal-2019,Fabri-etal-2021,Couto-Kowalewski-2022},
    photon up-conversion \cite{Gudem-Kowalewski-2022},
    polaritonic spectra \cite{Flick-Narang-2018,Fischer-Saalfrank-2021},
    photo-dissociation \cite{TorresSanchez-Feist-2021,Davidsson-Kowalewski-2020-jpc},
    and model refinements from the Cavity-Born-Oppenheimer approximation \cite{Flick-etal-jctc-2017,Schnappinger-Kowalewski-2023}.

    One of many challenges for this developing field is to build models
    that incorporate the details relevant for chemical reactivity
    (e.g.\ molecular vibrations along reaction coordinates)
    together with collective ensemble effects
    (several molecules coupled to the cavity field)
    \cite{Sidler-etal-2021,Schafer-2022}.
    Thus, in the last decade, several studies have turned their attention to such models of emitter ensembles \cite{Herrera-Spano-2016,Sidler-etal-2021,Sommer-etal-2021, Davidsson-Kowalewski-2020-jpc}.

\begin{figure}[t]
        \centering
        \includegraphics[width=\columnwidth]{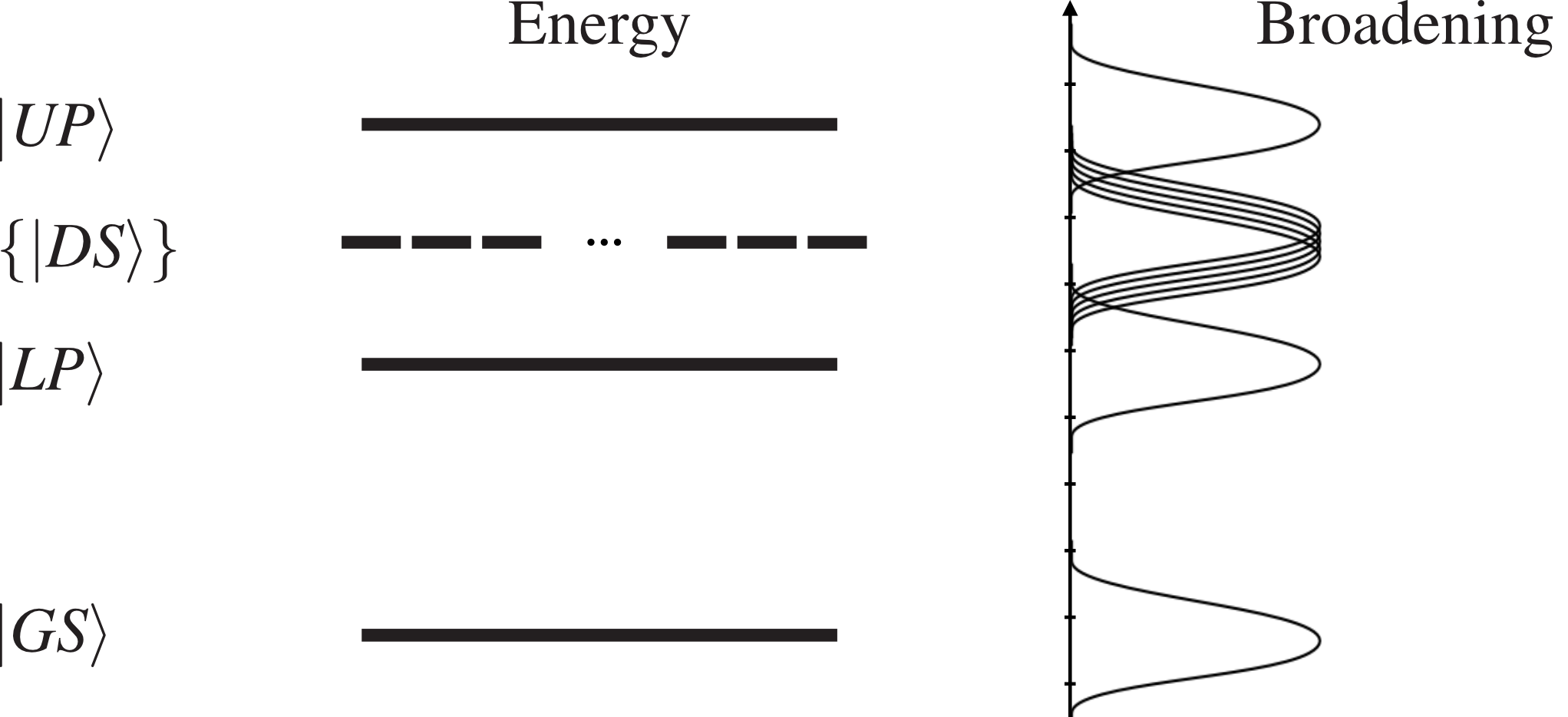}
        \caption{
            Schematic representation of the Tavis-Cummings model. The upper $\ke{UP}$
            and lower polartion state $\ke{LP}$ are split by a collective Rabi frequency.
            The $N-1$ dark states $\ke{DS}$ do not couple to the electromagnetic field and are
            thus unaffected by the cavity. Line broadening, cause by dephasing for example,
            introduces a coupling of the dark states with the polariton states.
        \label{fig:dark-states}}
    \end{figure}

    In these ensemble models, a common feature is the presence of dark states, which do not
    couple to the field of the cavity or an external field.
    These states are most straightforward to consider
    in a Tavis-Cummings model limited to a single excitation \cite{Tavis-Cummings-1968}:
    With $N$ identical emitters,
    there are two bright polaritonic states,
    $\ket{UP}$, and $\ket{LP}$,
    which are symmetrical under emitter permutation.
    The remaining $N-1$ polaritonic states are dark.
    See Fig.\ \ref{fig:dark-states}.
    In the Tavis-Cummings model, dark states are:
    degenerate \cite{Houdre-etal-1996},
    asymmetrical under emitter permutation
    \cite{Herrera-Spano-2017,Hernandez-Herrera-2019},
    have no transition dipole moment,
    and correspond to collective emitter excitations,
    which gain no population during Hamiltonian evolution \cite{Davidsson-Kowalewski-2020-jpc}.
    Note that these dark states emerge in the ensemble system,
    which is different from disallowed transitions in individual emitters.

    The absence of population in the dark states makes intuitive sense;
    assuming an initial symmetrical state,
    the Hamiltonian treats all emitters identically,
    so unitary evolution cannot induce asymmetry.
    The existence of the dark states, as well as their properties, are also quite robust;
    for example, they persist even if the field couplings vary among emitters
    \cite{Lopez-etal-2007}.
    In this work, the model will be an extended version of Tavis-Cummings model,
    but properties of the dark states still hold
    (degeneracy, asymmetry, non-existent transition dipole moment, and zero population under Hamiltonian evolution).

    Since the dark states are often assumed not to take part in the
    dynamics, they are commonly discarded them from the deployed models.
    However, there are numerous publications discussing how
    these states are physically significant.
    They have been studied in the broader context of cavity QED,
    \cite{Houdre-etal-1996,Yang-etal-1999,Zheng-2007,White-etal-2019,Botzung-etal-2020,Zanner-2022},
    quantum thermodynamics \cite{Scholes-etal-2020}, and polaritonic chemistry
    \cite{GonzalezBallestero-etal-2016,Sidler-etal-2021,Gera-Sebastian-2022,Cederbaum-2022}.
    In the latter context,
    these states are important both when the electromagnetic field couples to electronically excited states
    \cite{Vendrell-2018,Groenhof-etal-2019,Climent-etal-2022,Tichauer-etal-2022,Ulusoy-etal-2019},
    and vibrational states
    \cite{Pino-etal-2015,Herrera-Spano-2017,CamposGonzalezAngulo-etal-2019,Xiang-etal-2019,Du-YuenZhou-2022}.
    Despite the theoretical focus of numerous studies, the presence of dark states is by no means disregarded experimental studies \cite{Xiang-etal-2018,Eizner-etal-2019,Wersall-etal-2019,Liu-etal-2020,DelPo-etal-2020,Avramenko-Rury-2021,Cohn-etal-2022,Avramenko-Rury-2022}.

    The role of dephasing processes and its influence on dark-state dynamics
    is intertwined with identifying a mechanism for their population \cite{Tichauer-etal-2022}.
    To lift the decoupling of the dark states from the bright polariton states,
    an asymmetry in the Hamiltonian (or the Liouvilian) is required.
    To find such a process,
    we can consider processes that operate at the level of individual emitters,
    such as the interactions with each emitter's environment.
    We model this with dephasing operators,
    which can be included in the equations of motion by means of the Lindblad equation:
    \begin{equation}
        \label{eq:lindblad}
        \partial_t \hat \rho
        =
        -\frac{\i}{\hbar} \big[ \hat H \! , \hat \rho \big]
        +
        \sum_{n} \Big(
            \hat L_{n}\hat \rho \hat L^\dagger_{n}
            - \frac{1}{2}
            \big[ \hat L_{n}^\dagger \hat L_{n}, \hat \rho \big]_{+}
        \Big)\,.
    \end{equation}

    The emitter dephasing can be motivated by
    fluctuations in the relative energies between the states in the emitter systems
    \cite{Marquardt-Puttmann-2008},
    or by a build-up of system--environment correlations \cite{Costa-etal-2016}.

    In the context of strongly coupled molecular systems,
    there has been a recent focus on the utility of
    modelling decay processes with the Lindblad equation
    \cite{Davidsson-Kowalewski-2020-jcp,TorresSanchez-Feist-2021,Wellnitz-etal-2021}.
    A consequence of using a density matrix based time evolution,
    however, is an increase in computational complexity.
    For the situation where only decay processes out of the Hilbert space under consideration
    are involved, an effective non-Hermitian Hamiltonian together with a Schr\"odinger equation
    can be used \cite{Ulusoy-Vendrell-2020,Antoniou-etal-2020,Felicetti-etal-2020}.
    However, when dephasing becomes relevant, such simplifications will be harder to make.

    The quantum trajectories method (also called the stochastic Schrödinger equation \cite{Coccia-etal-2020} or Monte Carlo wave function \cite{Molmer-etal-1993} method) can be used
    to bypass the explicit evaluation of the density matrix.
    Here, we use this method to time-evolve the Lindblad equation
    using an ensemble of stochastic pure-state evolutions \cite{Dalibard-etal-1992,Molmer-etal-1993}.
    Quantum trajectories are not yet widely used for polaritonic chemistry problems,
    but some studies have started to emerge \cite{Triana-Herrera-2022}.

    In this work, we investigate the role of dephasing for molecular strong coupling
    and its effects on the dark state dynamics.
    We include the effects of both dephasing and photon decay.
    The model is based on the Lindblad equation and implemented with quantum trajectories.

    \section{System and model}

    The model system is based on an optical cavity with a single carbon monoxide molecule,
    where the fundamental cavity mode is resonant with a transition
    between the $^1\Sigma^+$ ground-state
    and an $^1\Pi$ electronically excited state in the CO molecule.
    To simulate a many particle system with dark states,
    which is also resonant with the cavity mode,
    $N$ two-level systems are added to the Hamiltonian.
    The Hamiltonian is thus an extension of the Tavis-Cummings model
    \cite{FriskKockum-etal-2019}
    (an optical cavity with $N$ two-level systems
    under the rotating wave approximation), where the
    extension is the the CO molecule with its internuclear coordinate.

    The Hamiltonian consists of the following five parts,
    \begin{equation}
        \label{eq:hamiltonian}
        \hat H
        =
        \hat H_{c}
        +
        \hat H_{a}
        +
        \hat H_{ca}
        +
        \hat H_{m}
        +
        \hat H_{cm}\,,
    \end{equation}
    where
    $\hat H_{c}$ corresponds to a single fundamental mode of linearly polarized light in an optical cavity,
    $\hat H_{a}$ are the $N$ two-level emitter systems (or atoms),
    $\hat H_{ca}$ are all the resonant cavity--atom interactions,
    $\hat H_{m}$ is the CO molecule,
    $\hat H_{cm}$ is the cavity--molecule interaction.

    \begin{figure}
        \centering
        \includegraphics[width=8.6cm]{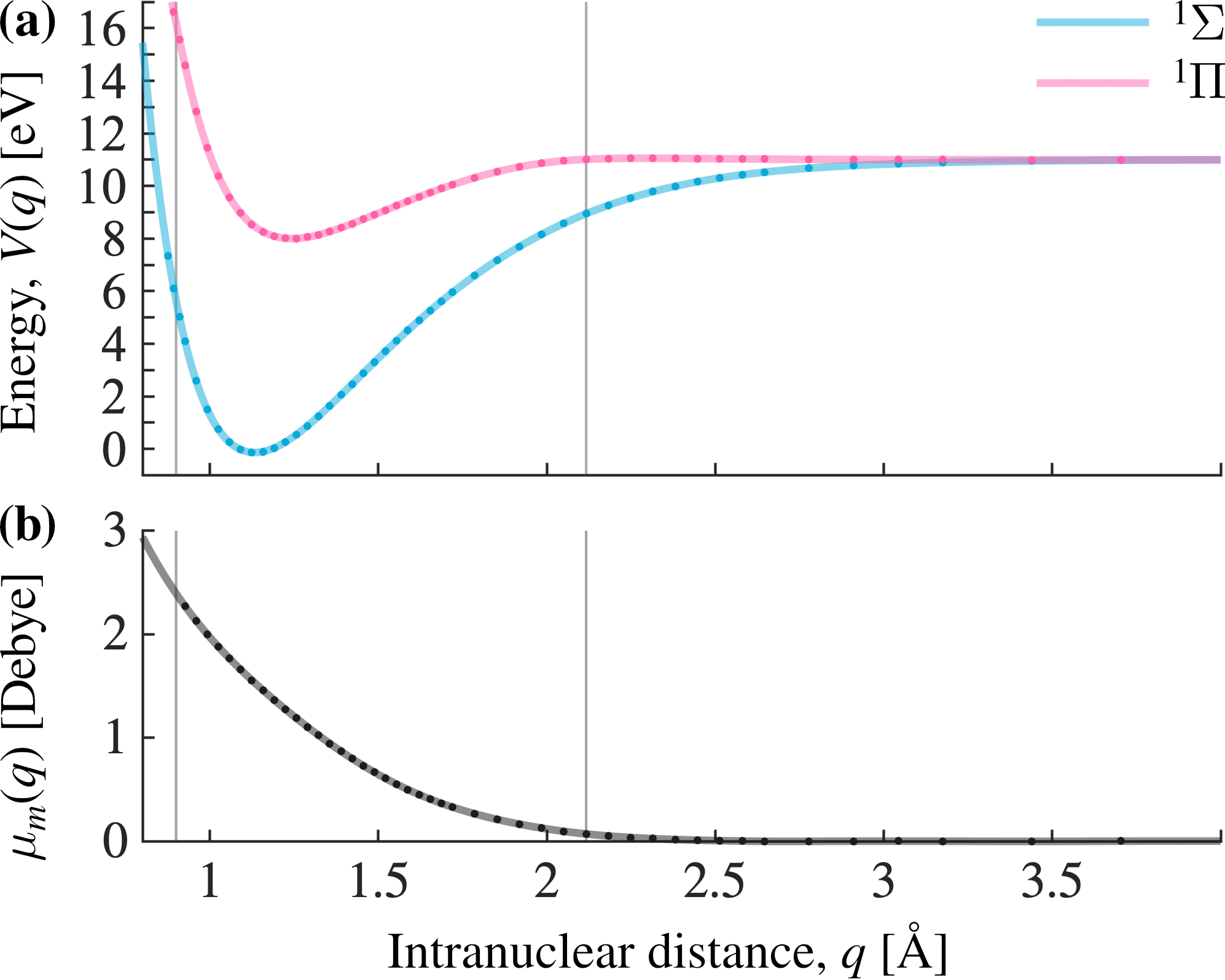}
        \caption{
            Potential energy curves and transition dipole curves for the CO molecule.
            Dots marks frozen internuclear distances from the quantum chemistry calculation.
            Vertical lines surround the region used in the time-evolution.
            (a) Potential energy surfaces for the $^1\Sigma$~ground-state and one $^1\Pi$ excited state.
            (b) Transition dipole moment between $^1\Sigma$~and $^1\Pi$.
            \label{fig:co}
        }
    \end{figure}

    The cavity mode is modelled in the Fock-basis
    $\{\ket{0}, \ket{1}, \ket{2}, \cdots\}$,
    and described by the following Hamiltonian:
    \begin{equation}
        \hat H_{c}
        =
        \hbar \omega_c \, \hat a^\dagger \! \hat a\,,
    \end{equation}
    where $\hat a^\dagger$ and $\hat a$ are the photon creation and annihilation operators respectively,
    and $\omega_c$ is the cavity mode frequency.
    The cavity photon energy is fixed at $\hbar \omega_c = 8.27\;$eV ($\lambda = 150\;$nm), % 8.2656
    which makes the transition between electronic states in the CO molecule resonant with the cavity.
    This choice puts the minima of the two potentials at similar energies,
    and prevents population from accumulating in either well,
    see Fig.\ \ref{fig:polaritonic-pes}.

    In addition to the CO molecule,
    the optical cavity contains $N$ two-level emitters (or atoms),
    where $N \in \{$0, 1, 2, 3, 5, 8, 13, 22, 36, 60$\}$ is a modulated parameter.
    \begin{equation}
        \hat H_{a}
        =
        \sum_{n=1}^N
        \hbar \omega_a \, \hat \sigma_{n}^\dagger \! \hat \sigma_{n}\,,
    \end{equation}
    where $\ket{g_n}$ and $\ket{e_n}$ are the ground and excited state
    of the $n$-th two-level system, respectively.
    The operators $\hat \sigma_{n} = \ket{g_n}\!\bra{e_n}$
    and $\hat \sigma_{n}^\dagger = \ket{e_n}\!\bra{g_n}$
    de-excite and excite the $n$-th two-level system, respectively.
    The energy of the transition is chosen to be resonant to the cavity's photon energy,
    $\hbar \omega_a = \hbar \omega_c = 8.27\;$eV. %8.2656

    The interaction between emitters and cavity mode is derived
    under the dipole and the rotating wave approximations:
    \begin{equation}
        \label{eq:cavity-atom-interaction}
        \hat H_{ca}
        =
        \sum_{n=1}^N
        \mathcal{E}_c(N) \, \mu_a
        \big( \hat a^\dagger \hat \sigma_{n} + \hat a \, \hat \sigma_{n}^\dagger \big)\,.
    \end{equation}
    The transition dipole moments of all two-level systems,
    $\mu_a$, are identical, and set to $1.5$\;Debye,
    which is on the same order of magnitude as the CO molecule
    (see Fig.\ \ref{fig:co}(b)).

    The CO molecule is modelled by including its electronic $^1\Sigma$ ground-state
    potential energy curve, and one of the two electronically excited $^1\Pi$ states,
    which has a non-zero transition dipole $\mu(q)$ moment perpendicular to the molecular axis,
    and is assumed to be aligned with the field polarisation.
    The molecular Hamiltonian reads:
    \begin{equation}
        \label{eq:ham-m}
        \hat H_{m}
        =
        - \frac{\hbar^2}{2M} \frac{\d^2}{\d q^2}
        + \hat V_\varSigma(q) \ket{\varSigma}\!\bra{\varSigma}
        + \hat V_\varPi(q) \ket{\varPi}\!\bra{\varPi}\,.
    \end{equation}
    Here, the first term is the kinetic energy of the nuclei, with
    the reduced mass of the CO molecule ($M=12\;498\;m_e$).
    The potential energy curves for $V_\varSigma(q)$ and $V_\varPi(q)$
    are shown in Fig.~\ref{fig:co}(a).

    The previous choice of photon energy makes the molecular transition
    resonant with the cavity mode at an internuclear distance of $q = 1.17$\;Å.
    \begin{equation}
        \label{eq:cavity-molecule-interaction}
        \hat H_{cm}
        =
        \mathcal{E}_c(N) \, \mu_{m}(q)
        \Big(
            \hat a^\dagger \ket{\varSigma}\!\bra{\varPi} + \hat a \ket{\varPi}\!\bra{\varSigma}
        \Big)\\[4pt]
    \end{equation}
    The transition dipole moment $\mu_{m}(\hat q)$ varies between about 0 and 3 Debye
    according to Fig.\ \ref{fig:co}(b).
    The vacuum electric field strength,
    $\mathcal{E}_c(N)$ is the same as the atoms are experiencing (see Eq.~\eqref{eq:cavity-atom-interaction}).

    The vacuum electric field strength, $\mathcal{E}_c(N)$,
    in Eqs.\ \eqref{eq:cavity-atom-interaction} and \eqref{eq:cavity-molecule-interaction},
    is fixed at $3.00\;\text{V}\!/\text{nm}$ for a single CO molecule model with no atoms (i.e. $N=0$).
    To compensate for an increase in collective coupling strength,
    as we increase the number of atoms in the model,
    the vacuum field strength is scaled by $1/\sqrt{N+1}$ \cite{Tichauer-etal-2022}.
    \begin{equation}
        \label{eq:coupling-strength}
        \mathcal{E}_c(N)
        =
        \frac{3.00}{\sqrt{N + 1}} \, \hspace{2pt}
        \text{V}\!/\text{nm}
    \end{equation}
    Note that the effects of single particle coupling strength can not be directly
    compared to a collective coupling strength of equal size. Thus, the
    results are expected to show a different behavior \cite{Davidsson-Kowalewski-2020-jpc}.

    The highest single particle cavity coupling strength,
    $g_\text{avg} = \mathcal{E}_c \mu_\text{avg}$,
    occurs for $N = 0$.
    We compare this to the typical energy scale,
    $g_\text{avg}/\hbar \omega_c = 0.011$.
    The system operates well below
    the ultra-strong coupling regime.
    Note that we assumed an average transition dipole moment of
    $\mu_\text{avg} = 1.5$\;Debye.

    In addition to the unitary part of the Hamiltonian,
    there are two non-unitary physical processes,
    which are introduced through the Lindblad equation \eqref{eq:lindblad}.
    The Lindblad framework assumes that,
    that the system-bath correlations times are short enough to allow
    for the Markovian approximation \cite{Manzano-2020,Pino-etal-2015}.
    Thus, our results address interactions that can be modelled as non-Markovian.

    The first non-unitary process is single-emitter dephasing
    of the CO molecule and all $N$ two-level systems.
    The corresponding rate of dephasing, $\gamma$,
    is the key parameter whose effects we aim to investigate.
    It is the same for all emitters,
    which have their individual dephasing operator
    \begin{equation}
        \label{eq:dephasing-op}
        \hat L_{n}
        \defeq
        \sqrt{\!\frac{\gamma}{2}} \;
        \hat \sigma_z^{(n)}\,,
    \end{equation}
    where, $\hat \sigma_z^{(n)}$ is the third Pauli matrix for the $n$-th emitter:
    \begin{equation}
        \hat \sigma_z
        =
        \begin{bmatrix}
            1 & 0 \\
            0 & -1
        \end{bmatrix}\,.
    \end{equation}
    Other choices of this operator yield the same time-evolution.
    However, this choice is beneficial for the
    quantum trajectory method, since it does not transfer population
    between states~\cite{Coccia-etal-2018}.

    When constructing the Lindblad dephasing operators,
    we neglect the fact that the sub-systems are coupled
    (by the cavity mode) and build them phenomenologically.
    This is generally considered acceptable outside the ultra-strong coupling regime \cite{Beaudoin-etal-2011,Salmon-etal-2022}.
    In the appendix, section \ref{sec:appendix-phenomenological},
    we discuss the potential problems with this approach,
    and argue that the phenomenological operators are appropriate.

    The second non-unitary process is photon decay caused by a lossy cavity.
    The Lindblad operator for a single cavity mode,
    with the photon decay rate $\kappa$, reads:
    \begin{equation}
        \label{eq:decay-op}
        \hat L_c
        \defeq
        \sqrt{\kappa} \, \hat a \, .
    \end{equation}
    The photon lifetime is fixed at $\tau = 100\;$fs,
    yielding $\kappa = 1/\tau = 0.01\;\text{fs}^{-1}$.
    For the chosen photon energy at $\hbar\omega_c=8.27\;$eV,
    this corresponds to a quality factor of $Q = 1.26\times 10^{3}$.
    For comparison, Q-factors reported in the literature
    range from
    $10^1$ \cite{Wersall-etal-2019,Franke-etal-2020}
    for plasmonic nanoparticles \cite{Chikkaraddy-2016}
    to $10^4$ \cite{Najer-etal-2019,Wellnitz-etal-2022,Franke-etal-2020}
    for Fabry--Pérot cavities made of Bragg reflectors.

    \section{Methods}

    Our studied observable is energy retention,
    i.e.\ we consider the fraction of the initial excitation that,
    despite the photon decay,
    remains in the system after 500\;fs.
    The duration is chosen in relation to the timescale of the nuclear dynamics
    and mirrors previous investigations
    \cite{Davidsson-Kowalewski-2020-jpc,Davidsson-Kowalewski-2020-jcp}.
    The initial excitation consists of the cavity mode being is in its first excited (single photon) state,
    while all emitters are in their ground-states.
    This limits the total number of excitations to one,
    which allows us to truncate the basis.

    A direct solution of the Lindblad equation \eqref{eq:lindblad}
    with a density matrix scales quadratically
    with the number of states involved.
    Such an approach becomes prohibitive as the number of atoms in our system increases.
    Instead of modelling a statistical state as a single density operator,
    we use quantum trajectories \cite{Dalibard-etal-1992,Molmer-etal-1993} to obtain the statistics from an ensemble of pure state wave functions.
    Thus, the cost is shifted from a single memory consuming density matrix,
    to running multiple, but lighter, pure-state calculations with wave functions, $\{\psi_i(t)\}$.
    Each wave function evolves stochastically,
    with "quantum jumps" occurring randomly in proportion to physical parameters.
    One can prove that
    in the limit of an infinite ensemble of wave functions,
    the state from evolution with quantum trajectories
    approaches the state from the Lindblad equation
    \cite{Molmer-etal-1993}.

    From the ensemble of $N_{T}$ trajectories, $\{\psi_i(t):i\in 1\cdots N_{T}\}$,
    a density matrix can be recovered by summing outer products of each wave function with a uniform weight:
    \begin{equation}
        \label{eq:statistical-state-rho}
        \hat \rho
        =
        \frac{1}{N_{T}}
        \sum_{i=1}^{N_{T}} \ket{\psi_i} \! \bra{\psi_i}\,.
    \end{equation}
    Expectation values $\{\braket{\hat A_i}\}$ can also be obtained
    directly from the weighted sum of all trajectories,
    without having to construct $\hat \rho$ explicitly:
    \begin{equation}
        \text{Tr}\big[\hat A \hat \rho \big]
        =
        \frac{1}{N_{T}}
        \sum_{i=1}^{N_{T}} \braket{\hat A_i}\,.
    \end{equation}

    The quantum trajectory method requires two modifications to
    the time-dependent Schr\"odinger equation.
    The first one adds the last term from Eq.~\eqref{eq:lindblad},
    i.e.\ $\sum_n -1/2 \, [ \hat L_{n}^\dagger \hat L_{n}, \hat \rho ]_{+}$,
    to the Hamiltonian in the form of a norm-decaying term.
    However, in our choice of implementation, the wave function is continuously renormalized
    (at each discrete time-step).
    Using the Lindblad operators from Eqs.\ \eqref{eq:dephasing-op} and \eqref{eq:decay-op}
    leads to the non-Hermitian Hamiltonian:
    \begin{equation}
        \label{eq:non-herm-ham}
        \hat H'
        \defeq
        \hat H
        -
        \i\hbar
        \frac{\kappa}{2} \,
        \hat a^\dagger \! \hat a
        -
        \i\hbar \frac{\gamma}{2}
        \sum_{n}
        \big(\hat \sigma_z^{(n)} \big)^{\!\dagger}
        \hat \sigma_z^{(n)}\,.
    \end{equation}
    The second term in Eq.\ \eqref{eq:non-herm-ham}, is responsible for
    the photon decay, while the third term only affects the norm of the total wave function,
    since $\big(\hat \sigma_z^{(n)} \big)^{\dagger} \hat \sigma_z^{(n)} = \hat \1$.
    The algorithm used for the propagation renormalizes
    the wave function in each step,
    thus the last term from Eq.\ \eqref{eq:non-herm-ham} has no effect and can be removed.

    The second part that is required for in the quantum trajectories method are discrete,
    stochastic jumps.
    These jumps originate from the first term in Eq.\ \eqref{eq:lindblad},
    i.e.\ $\sum_n \hat L_{n}\hat \rho \hat L^\dagger_{n}$.
    The probability of a jump occurring during a time-interval $\Delta \.t$
    depends on the rates $\kappa$ and $\gamma$ as well as the population in the subspace from which the jump occurs:
    \begin{equation}
        P\big(\.\text{jump with }\hat L_k \text{ during } \Delta t \big)
        =
        \Delta \. t \braket{\varPsi(t) |\hat L^\dagger_k \hat L_k | \varPsi(t)}\,.
    \end{equation}
    Note here that the Lindblad operators $L_k$ include the rates, $\kappa$ and $\gamma$,
    as shown in Eqs.\ \eqref{fig:deph-op} and \eqref{eq:decay-op}.
    At each time-step, random numbers are generated to determine if a jump occurs,
    in which case the Lindblad operator $\hat L_k$, is applied to the wave function
    and the wave function is normalized.

\begin{figure*}[t]
        \centering
        \includegraphics[width=\textwidth]{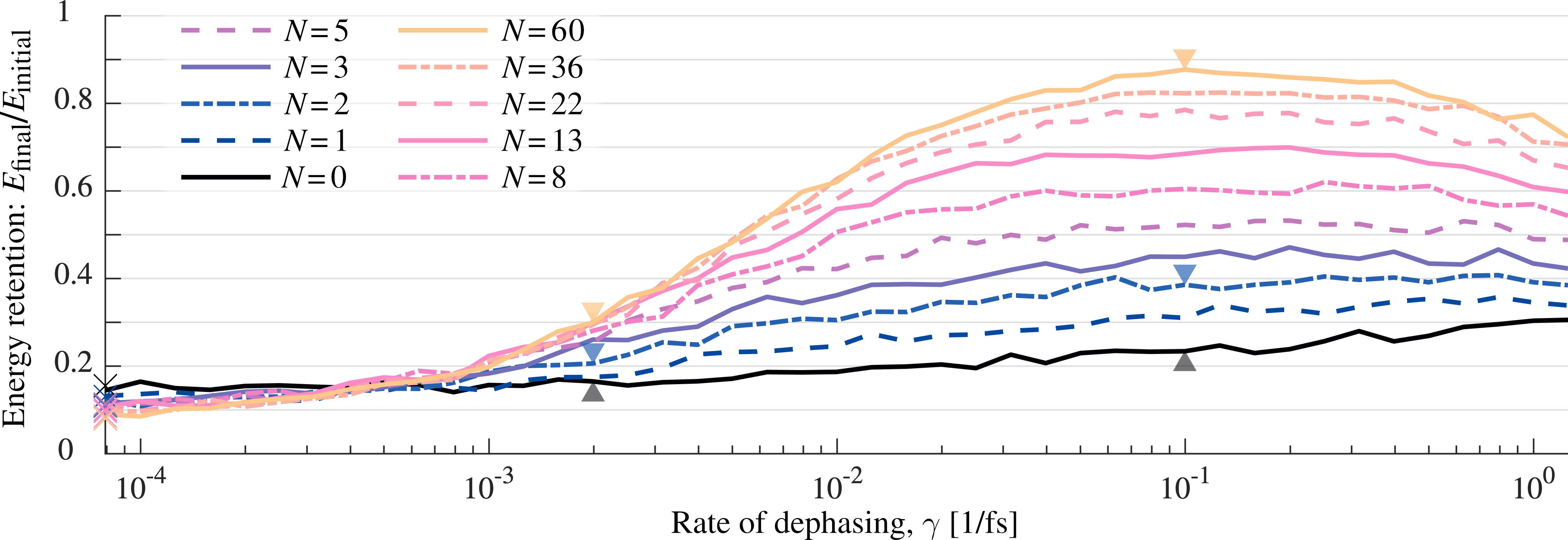}
        \caption{
            Energy retention in relation to dephasing rate.
            The vertical axis shows energy retention,
            i.e.\ the remaining energy as a fraction of initial energy,
            $E_\text{final} / E_\text{initial}$ where $E_\text{initial} = \hbar \omega_c$,
            after 500\;fs of time-evolution.
            The horizontal axis is the single emitter dephasing rate,
            $\gamma$, which is the same for all emitters.
            Each curve corresponds to a particular number of two-level systems, $N$.
            The number of dark states for each curve is $N-1$.
            Crosses on the vertical axis shows the energy retention for $\gamma = 0$.
            Arrows mark points where dynamics plots are supplied,
            see Figs.\ \ref{fig:population-n-0}, \ref{fig:population-n-2}, and \ref{fig:population-n-60}.
            Standard deviations due to the Trajectories method (from a set of 25 runs) are also calculated at these points.
            Grey arrows both have $\sigma \approx 0.008$, % leeft: 0.007907840326, right: 0.008075558608
            left blue has $\sigma \approx 0.011$, % 0.01108772121
            right blue has $\sigma \approx 0.012$, % 0.01227856573
            left yellow has $\sigma \approx 0.009$, % 0.008778223607
            right yellow has $\sigma \approx 0.008$. % 0.007946963775
            \label{fig:data}
        }
    \end{figure*}

    We have implemented the quantum trajectories approach with our in-house software package QDng,
    which allows for the time evolution of wave functions.
    Implementations into existing methods are recurring in the literature
    \cite{Coccia-etal-2018,Mandal-etal-2022,Triana-Herrera-2022}.
    For each statistical state (see Eq. \eqref{eq:statistical-state-rho}),
    $N_{T} = 2500$ wave function trajectories were run.
    Estimates of the resulting errors are given in the caption of Fig.~\ref{fig:data}.
    Each wave function was time-evolved with the Arnoldi propagation method \cite{Smyth98cpc},
    at order 10 and a time-step of to $0.5 \. \text{au}$ ($0.012 \; \text{fs}$).

    The time-evolution is carried out in a product basis composed
    of the field-free molecular states, the field free two-level system states,
    and the Fock states of the cavity mode.
    In the following, we will refer to this basis as product basis.

    \begin{figure}[t]
        \centering
        \includegraphics[width=8.6cm]{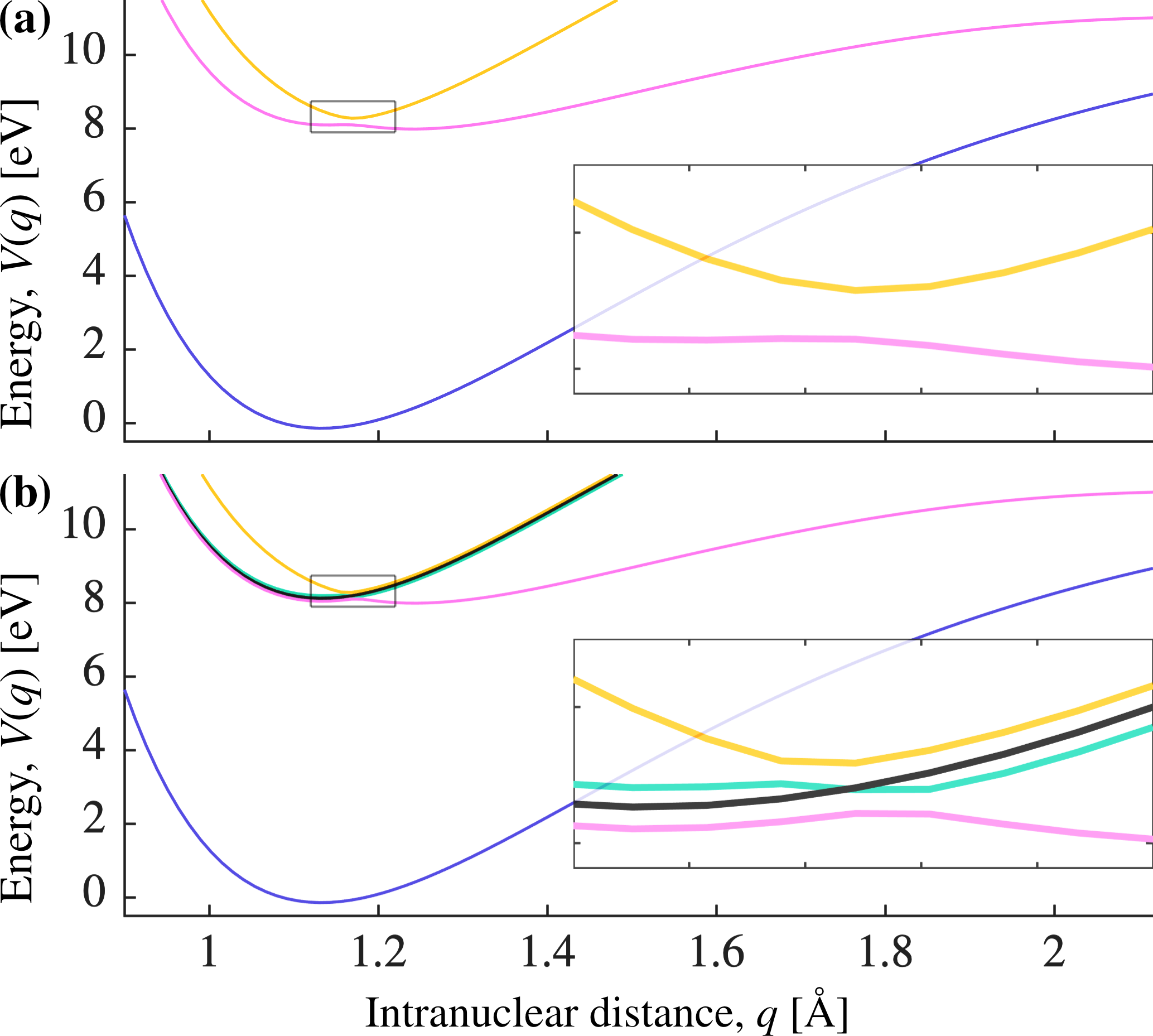}
        \caption{
            Potential energy surfaces of the polaritonic basis.
            A single CO molecule and $N = 2$ two-level systems in an optical cavity.
            With $N = 2$ the first dark state is introduced.
            The upper polaritonic state is yellow,
            the middle polaritonic state is green,
            the lower polaritonic state is pink,
            the ground-state is blue,
            and dark states are black.
            For higher values of $N > 2$
            the curves are essentially the same \cite{Davidsson-Kowalewski-2020-jpc},
            but additional degenerate dark states are introduced.
            \label{fig:polaritonic-pes}
        }
    \end{figure}

    \begin{figure}[t]
        \centering
        \includegraphics[width=8.6cm]{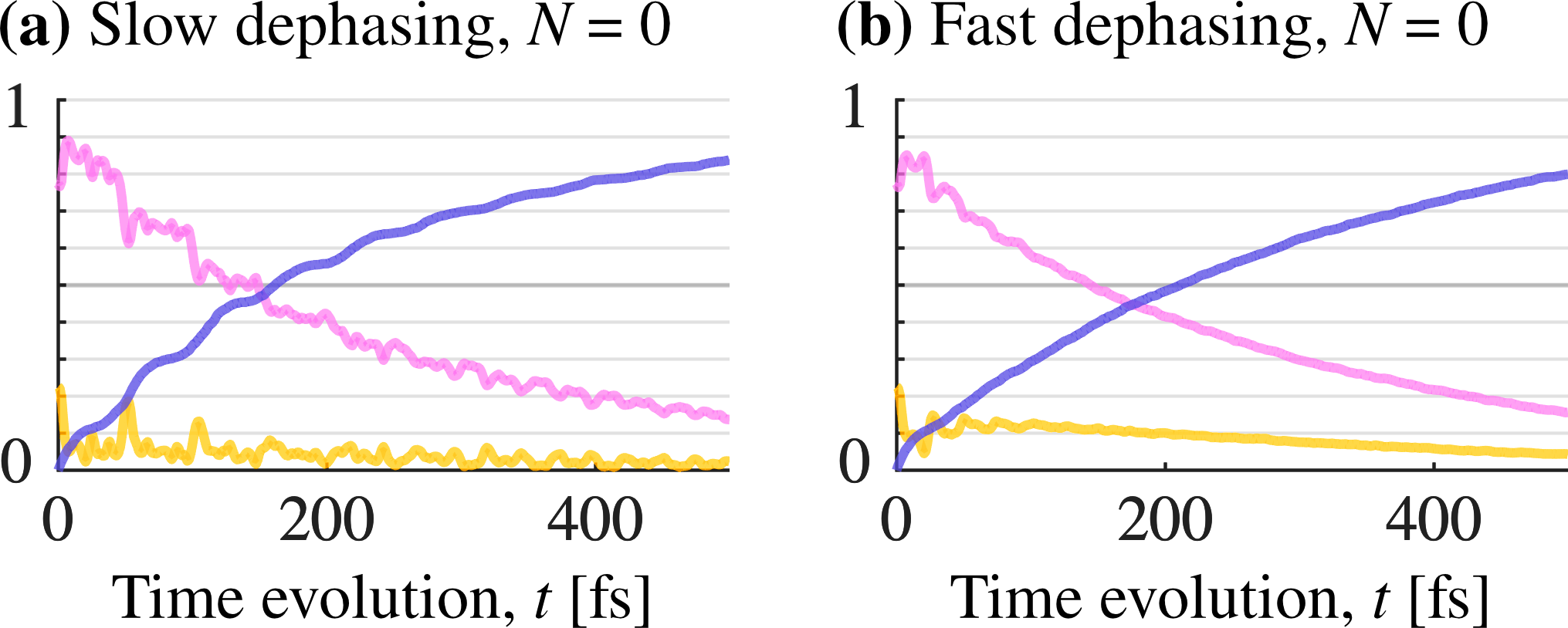}
        \caption{
            Population time-evolution of the polaritonic potential energy surfaces shown in
            Fig.\ \ref{fig:polaritonic-pes}(a).
            Colors match the ones used for the surfaces in
            Fig.\ \ref{fig:polaritonic-pes}.
            In the left column (a) the dephasing is slow
            ($\gamma \approx 0.002 \;\text{fs}^{-1}$),
            in the right column (b) dephasing is fast
            ($\gamma \approx 0.09 \;\text{fs}^{-1}$).
            The left grey arrows in Fig.\ \ref{fig:data} shows the data point of (a),
            and the right grey arrow is the data point for (b).
            The two cases, (a) and (b), behave very similar,
            the most apparent difference is the dampening of interference for fast dephasing,
            which makes the curves in (b) smoother.
            \label{fig:population-n-0}
        }
    \end{figure}

    For the purpose of interpreting the data (section \ref{sec:results})
    we diagonalize the potential curves,
    to obtain polaritonic curves (see Fig.\ \ref{fig:polaritonic-pes}).
    This can be achieved by a pointwise diagonalization of the Hamiltonian
    along the reaction coordinate, $q$,
    while omitting the kinetic energy operator.
    The number of polaritonic energy surfaces depends on $N$,
    but for $N > 2$ all additional energy surfaces are the degenerate dark states.
    Thus, plotting the result for $N = 0$ and $N = 2$ gives an overview for all other values of $N$
    (see Fig.\ \ref{fig:polaritonic-pes}(b)).
    Note that dark states are a feature of the polaritonic basis,
    and does not appear in the product basis.
    We ensure that uncoupled polaritonic surfaces are allowed to cross,
    by following the eigenvector tracking method described in \cite{Davidsson-Kowalewski-2020-jpc}.

    The potential energy curves and dipole functions of the CO molecule,
    shown in Fig.\ \ref{fig:co},
    are the relevant results of a quantum chemistry calculation
    that originally included eight non-degenerate states
    and their respective transition dipole moments.
    The electronic structure calculations were carried
    out with the program package Molpro \cite{Molpro-1,Molpro-2,Molpro-3} at  the
    CASSCF(10/14)/MRCI/aug-cc-pVQZ level of theory,
    with a state average over a total of twelve electronic states.
    Energies, dipole moments, and transition dipole moments are calculated at
    50 internuclear distances, between 0.926\;Å and 6.35\;Å.
    The two electronic states included in this work are
    $^1\Sigma^+$ and one of the two doubly degenerate $^1\Pi$ states.
    The data is in good agreement with previous calculations \cite{Shi-etal-2013,ONeil-Schaefer-1970}.
    The molecular potentials and transitions moments are
    interpolated to a spatial grid with 96 grid points in
    the interval $0.90 \leq q \leq 2.12$\,Å
    (see vertical lines in Fig.\ \ref{fig:co}).
    The result is used to construct the molecular Hamiltonians,
    $\hat H_{m}$ and~$\hat H_{cm}$ in Eqs. \eqref{eq:ham-m} and \eqref{eq:cavity-molecule-interaction},
    for the numerical calculations.

    \section{Results and Discussion}
    \label{sec:results}

    To construct a molecular Tavis-Cummings model, which includes dark states,
    $N$ two-level systems were introduced along with the CO molecule.
    We included the values $N \in \{$0, 1, 2, 3, 5, 8, 13, 22, 36, 60$\}$
    and for each value the dephasing rate is varied between 0
    and 10\% of the cavity frequency $\omega_c$,
    which corresponds to the dephasing rates: \ $0 \leq \gamma \leq 1.26 \; \text{fs}^{-1}$.
    After 500\;fs of time evolution, the energy retention is recorded.
    The remaining energy in the system is then plotted as a fraction of the initial energy,
    in relation to the dephasing rate for each $N$.
    This constitutes the main result in this study and is shown in Fig.~\ref{fig:data}.
    Note that the obtained curves are not perfectly smooth, which is an
    expected result of the stochastic sampling of the density matrix.

    Only 10\% to 15\% of the initial energy is retained in the system
    for the dephasing free case
    ($\gamma=0$, see crosses on the vertical axis in Fig.\ \ref{fig:data}).
    With no dephasing, the $N - 1$ dark states are not populated
    and does not impact the behavior of the system.
    However, even though the number of bright states are fixed,
    and with a constant collective coupling strength
    (see Eq.\ \eqref{eq:coupling-strength}),
    the energy retention for $\gamma=0$ varies.
    This can explained as follows: the collective Rabi splitting appears to behave
    similar to the single particle Rabi splitting between the upper and lower polariton.
    However, the splitting between middle polariton state and the upper and lower polariton
    state follows the single particle Rabi splitting and not the collective Rabi splitting  \cite{Davidsson-Kowalewski-2020-jpc}.

    With no two-level systems, i.e.\ $N = 0$,
    the impact of dephasing on the energy retention is small (black curve in Fig.~\ref{fig:data}).
    Fig.~\ref{fig:population-n-0}(a) and (b),
    compare the populations in the polaritonic basis
    (from Fig.\ \ref{fig:polaritonic-pes}(a)) for a slow and fast dephasing rate (and $N = 0$).
    The most obvious difference in the populations is the amount of oscillations caused by interference and by the avoided crossing in polaritonic states.
    We can understand the dampened oscillations as dephasing canceling
    the otherwise coherent population transfer between different trajectories.

    \begin{figure}[t]
        \centering
        \includegraphics[width=8.6cm]{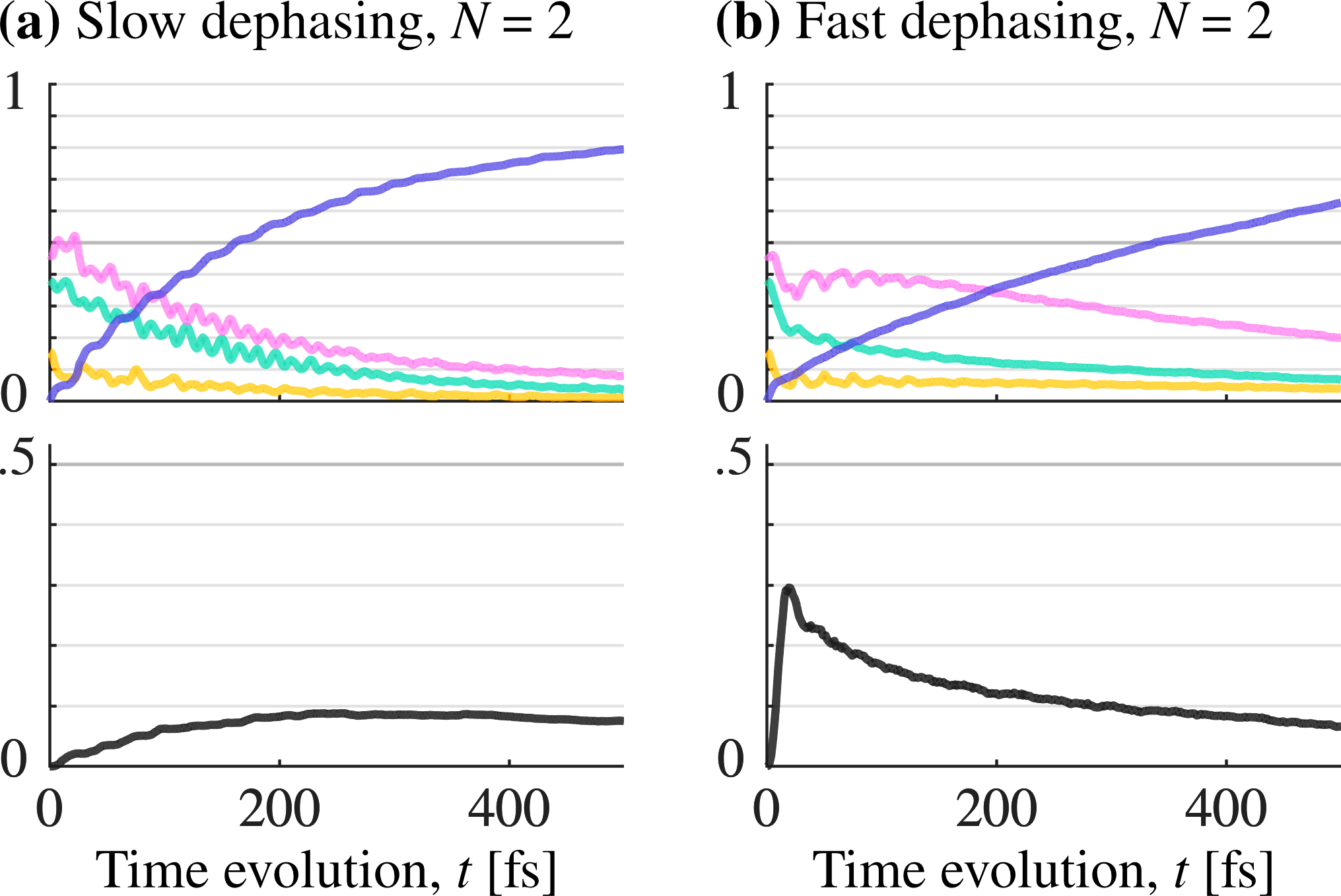}
        \caption{
            Population time-evolution of the polaritonic potential energy surfaces shown in
            Fig.\ \ref{fig:polaritonic-pes}(b).
            Colors match the ones used for the surfaces in
            Fig.\ \ref{fig:polaritonic-pes}.
            In the left column (a) the dephasing is slow
            ($\gamma \approx 0.002 \,\text{fs}^{-1}$),
            in the right column (b) dephasing is fast
            ($\gamma \approx 0.09 \,\text{fs}^{-1}$).
            The left blue arrows in Fig.\ \ref{fig:data} shows the data point of (a),
            and the blue arrow is the data point for (b).
            \label{fig:population-n-2}
        }
    \end{figure}

    Introducing dark states ($N \geq 2$) will make the energy retention
    increasingly sensitive to changes in the dephasing rate.
    We consider first the case with a single dark state ($N = 2$).
    Here, dephasing has an observable effect on energy retention.
    Fig.\ \ref{fig:polaritonic-pes}(b) shows the polaritonic energy surfaces for $N = 2$,
    which includes a single dark state (black curve) and middle polariton state (green curve).
    The corresponding population evolution
    is shown in Fig.\ \ref{fig:population-n-2},
    for slow dephasing (a) and fast dephasing (b).
    Note, how a fast dephasing rate imposes a rapid build-up of population in the dark state, which peaks at about 20\;fs and slowly decays thereafter.
    This in contrast to the slow dephasing rate, where the dark state population occurs
    slowly over a timescale of $\approx 300$\,fs.

\begin{figure}[t]
        \centering
        \includegraphics[width=8.6cm]{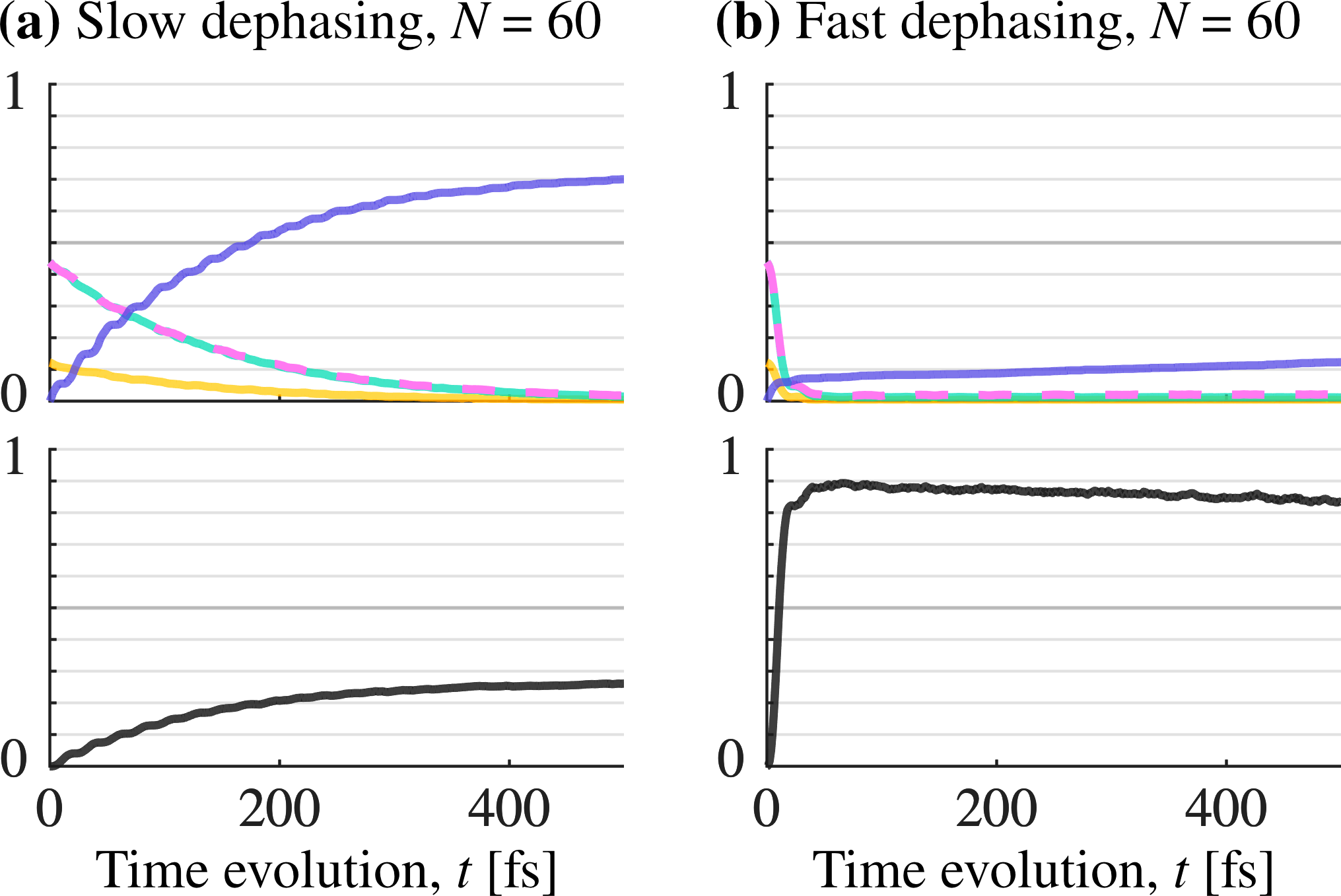}
        \caption{
            Population time-evolution of the polaritonic potential energy surfaces for $N = 60$
            (which are very similar to the ones shown in
            Fig.~\ref{fig:polaritonic-pes}.
            Colors match the ones used for the surfaces in
            Fig.\ \ref{fig:polaritonic-pes}(b),
            but the black curve is a sum of all 59 dark states.
            In the left column (a) the dephasing is slow
            ($\gamma \approx 0.002 \;\text{fs}^{-1}$),
            in the right column (b) dephasing is fast
            ($\gamma \approx 0.09 \;\text{fs}^{-1}$).
            The left yellow arrows in Fig.\ \ref{fig:data} shows the data point of (a),
            and the right yellow arrow is the data point for (b).
            \label{fig:population-n-60}
        }
    \end{figure}

    The highest energy retention is observed for
    the largest number of two-level system studied ($N = 60$)
    and a dephasing rate on the order of $\gamma \approx 10^{-1}\,\text{fs}^{-1}$.
    Here, the solid yellow curve in Fig.\ \ref{fig:data}
    shows a sharp increase in energy retention with increasing dephasing,
    from around 10\,\% to almost 90\,\%.
    The time-evolution of the populations for $N = 60$ is shown in Fig.\ \ref{fig:population-n-60},
    for both slow dephasing in (a) and fast dephasing in (b).
    The populations of all 59 dark states are shown as a sum (black curve).
    For the fast dephasing rate,
    the dark states population builds up rapidly,
    and reaches its maximum value of 90\,\% in about 40\,fs.
    Thereafter, the dark states population is very slowly released back to the bright states,
    with a retention of 80\,\% at 500\,fs.
    For $N = 60$ and a slow dephasing rate,
    the build up to about 27\,\% is slow,
    and does not reach a maximum within 500\,fs.
    For both slow and fast dephasing,
    the population in the dark states has significantly increased
    when compared to the case $N = 2$
    (compare Figs.\ \ref{fig:population-n-2} and \ref{fig:population-n-60}).

    The mechanism for the energy retention can be explained by
    the transition dipole moments of the collective system.
    The main energy loss mechanism is the decay of photons due to imperfect cavity mirrors.
    However, the dark states have no transition dipole moment in the absence of dephasing,
    as they represent asymmetric superpositions of matter excitations.
    Thus, the dark states do not couple to the cavity mode in an ideal system
    and can not scatter photons into the cavity mode.
    Population in the dark states are thus protected from photon
    decay. However, the dephasing breaks the symmetry in the Hamiltonian
    that decouples the dark states from the upper, middle, and lower,
    polariton states and enables population transfer.
    With an increasing number of particles $N$, the number of
    available dark states increases, and thus the effective rate
    increases with which these states are populated increases.
    The dark state subspace can thus serve as a reservoir which
    protects the systems from photon decay \cite{Pino-etal-2015,Quach-etal-2022}.

    The population transfer between dark states and bright states goes both ways.
    As can be seen in Figs.\ \ref{fig:population-n-2}(b)
    and \ref{fig:population-n-60}(b), dephasing will also slowly return
    the population to the bright states. However, this process
    is slower and than the population transfer into the dark states,
    resulting in an overall slow down of the energy loss.

    The energy retention in Fig.\ \ref{fig:data} has a local maximum with respect
    to the dephasing rate at about $\gamma \approx 10^{-1}$\,fs$^{-1}$ and $N=60$.
    In this regime, the Rabi oscillations ($\Omega_R = 2.85\times10^{-1}\,\text{fs}^{-1}$) are dampened, but not yet over dampened. The line width of the dark states and the polariton states are sufficiently narrow to be spectrally well separated.

    However, if the dephasing
    rate the is further increased, the polariton states begin to overlap with the dark states.
    In this over dampened regime, the dark states are no longer sufficiently decoupled from the
    bright states, and there is a significant leakage from the dark states.
    This effect causes a decreased energy retention for large dephasing rates.
    Note that the maximum of the energy retention shifts to larger dephasing rates
    with decreasing $N$.
    Note that the decrease in energy retention, after the maximum,
    depends on the choice of our initial state (excited cavity),
    through the momentary rates populating and depopulating the
    dark state reservoir.

    \section{Conclusion}
    In summary, we have investigated a molecular Tavis--Cummings model
    under the influence of dephasing and photon decay.
    The model system consisted of a CO molecule with a varying number
    of resonantly coupled two-level systems, which has been
    solved with a quantum trajectory approach instead of a
    direct solution of the Lindblad equation. This approach allowed
    us to use a wave function base calculation which scales more
    favorably with respect to the number of states than evaluating
    a density matrix explicitly.

    In this atomistic model, we could show that the dark states become increasingly
    coupled to the polariton states as the dephasing rate is increased.
    As a result, population gets trapped in the dark states, and
    it is protected from photon decay processes. Our findings
    are in line with earlier studies \cite{Pino-etal-2015,Quach-etal-2022}.
    Under the influence of dephasing the dark states are no
    longer decoupled from the polariton states. The effective
    transfer rate into the dark states even increases with an increasing number of dark
    states, and thus providing an increasingly efficient protection against photon decay.
    In the investigated systems, the photonic excitation was rapidly transferred
    into the dark state reservoir and slowly released back into the bright polariton states.

    Our results show that dephasing, as it would occur under experimental conditions
    in condensed phase, plays a significant role in the dynamics of such a system.
    Realistic models should thus not only include photon decay, which has been demonstrated
    to play a crucial role \cite{Davidsson-Kowalewski-2020-jpc},
    but should also include the effects from pure dephasing in condensed phase.

    \section{Acknowledgments}

    \label{sec:acknowledgments}
    This project has received funding from the European Research Council (ERC)
    under the European Union’s Horizon 2020 research and innovation program (grant agreement No. 852286) and European Union's Horizon 2020 research and innovation program under the Marie
    Sklodowska-Curie grant agreement no.
    E.D. thanks Prof.\ Jonas Larson for valuable discussions and insightful feedback.

    \section{Data availability}

    The data that support the findings of this study are available upon reasonable request.

    \section{Appendix}

    \begin{figure}[b]
        \centering
        \includegraphics[width=\columnwidth]{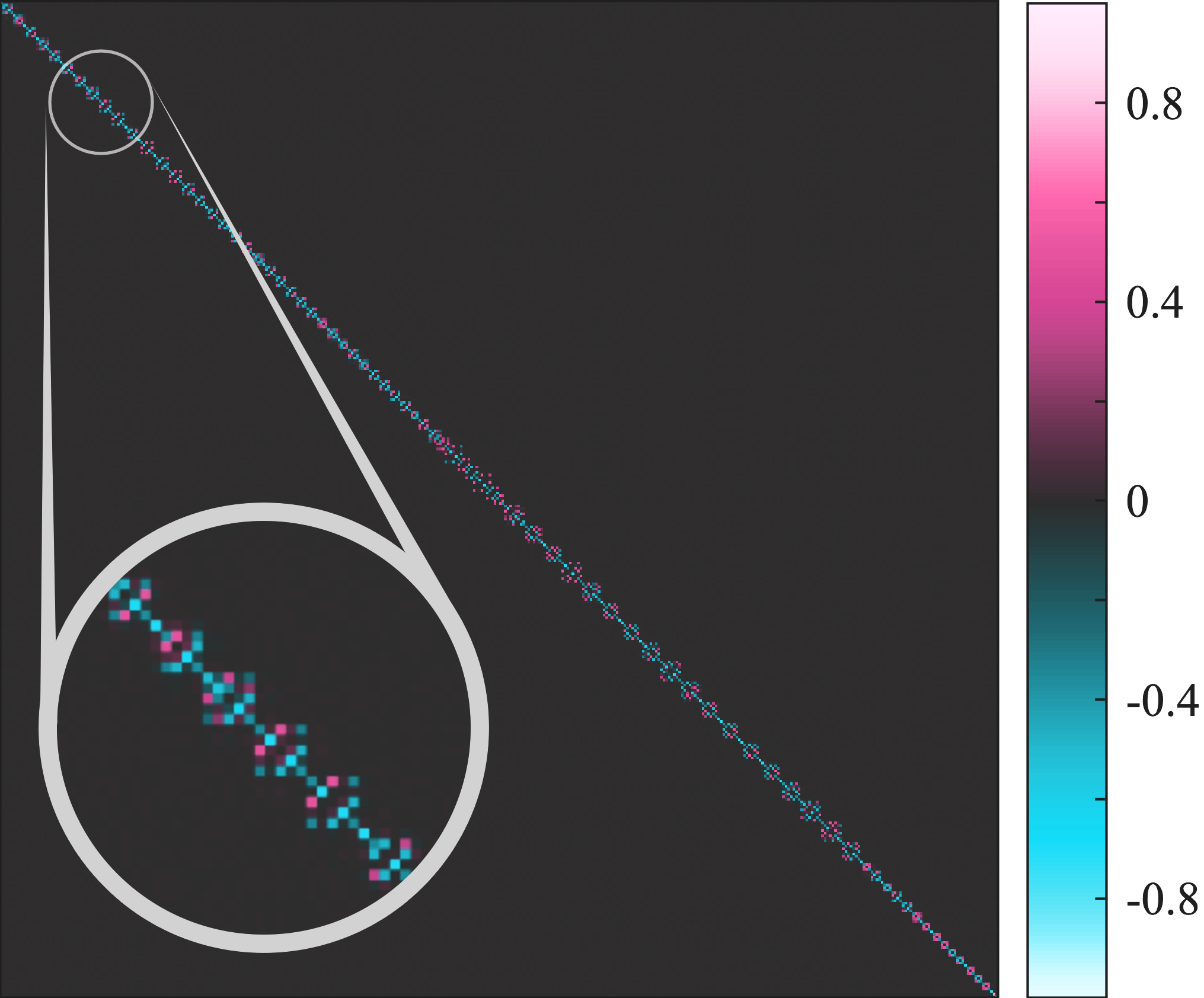}
        \caption{Dephasing operator after transformation to eigen-basis.
        The initially diagonal operator acquires off-diagonal elements,
        but the operator remains essentially block-diagonal.
        \label{fig:deph-op}}
    \end{figure}

    \label{sec:appendix}
    \subsection{Impacts from phenomenological dephasing operators}
    \label{sec:appendix-phenomenological}
    For a single two-level system,
    the dephasing operator in Eq.\ \eqref{eq:dephasing-op} has the expected effect;
    to decay away phase information without moving population between states.
    However, when a system is built from many strongly coupled sub-systems (such as in our case),
    this operator will include some unintended amount of driving and decay.
    To probe how significant this effect is,
    we can fully diagonalise the Hamiltonian from Eq.\ \eqref{eq:hamiltonian},
    and apply the basis transformation to a dephasing operator.
    (Note that we can omit the ground-state from the diagonalisation
    since it does not couple to any other state.)
    This basis transformation into a fully diagonalised Hamiltonian will give a set of eigenstates containing some unphysical states;
    when the energy of the vibrational motion is enough to dissociate the molecule.
    However, since our time-evolution does not exhibit dissociation we do not populate such questionable states,
    and they could be excluded from this analysis without affecting the conclusion.

    A dephasing operator (for $N = 2$ atoms) after transformation
    to the energy ordered eigenbasis is shown in Fig.\ \ref{fig:deph-op}.
    Driving and decay will appear as off-diagonal elements.

    On close inspection we can see that there are off-diagonal elements,
    but that the matrix is essentially block diagonal.
    This means that there are some driving and decay between eigenstates that are close to degeneracy,
    but the population is trapped within each block of states.
    Thus, we cannot drive or decay population further than
    between a few energetically adjacent states.
    This is not of concern to our investigation
    and a phenomenological operator is considered sufficient.
    For further discussions, see for instance
    \cite{Scala-etal-2007,Betzholz-etal-2021,Pino-etal-2015}.

    \bibliography{Bibliography}

\end{document}